\documentclass[aps,prx,twocolumn,nofootinbib,superscriptaddress,showpacs,showkeys,longbibliography]{revtex4-2}
\usepackage{graphicx}
\usepackage{amsmath, amsxtra, amssymb, mathtools, isomath, nccmath, xspace, siunitx}
\usepackage{color}
\usepackage[utf8]{inputenc}
\usepackage[british]{babel}
\usepackage{hyperref}
\usepackage{sidecap}
\usepackage{natbib}
\usepackage{dsfont}
\usepackage{xcolor}
\usepackage{pifont}
\newcommand\T{\rule{0pt}{2.7ex}}       
\newcommand\I{\rule{0pt}{2.2ex}} 	   
\newcommand\B{\rule[-1.8ex]{0pt}{0pt}} 

\renewcommand{\vec}{\vectorsym}

\newcommand{\ket}[1]{\ensuremath{\lvert #1 \rangle}\xspace}%

\renewcommand{\theequation}{A\arabic{equation}}
\renewcommand{\vec}{\vectorsym}
\sidecaptionvpos{figure}{h}
\long\def\symbolfootnote[#1]#2{\begingroup%
\def\thefootnote{\fnsymbol{footnote}}\footnotetext[#1]{#2}\endgroup}

\newcommand{\Aref}[1]{\hyperref[#1]{A}}
\newcommand{\Bref}[1]{\hyperref[#1]{B}}
\newcommand{\Cref}[1]{\hyperref[#1]{C}}
\newcommand{\Dref}[1]{\hyperref[#1]{D}}
\newcounter{myequation}
\makeatletter
\@addtoreset{equation}{myequation}
\makeatother

\begin{document}
\setlength{\belowdisplayskip}{9pt}
\setlength{\abovedisplayskip}{9pt}

\title{\bf{Microscopic electronic structure tomography of Rydberg macrodimers}}


\author{Simon Hollerith}
\affiliation{Max-Planck-Institut f\"{u}r Quantenoptik, 85748 Garching, Germany}
\affiliation{Munich Center for Quantum Science and Technology (MCQST), 80799 Munich, Germany}

\author{Jun Rui}
\email[]{Jun.Rui@mpq.mpg.de}
\affiliation{Max-Planck-Institut f\"{u}r Quantenoptik, 85748 Garching, Germany}
\affiliation{Munich Center for Quantum Science and Technology (MCQST), 80799 Munich, Germany}

\author{Antonio Rubio-Abadal}
\affiliation{Max-Planck-Institut f\"{u}r Quantenoptik, 85748 Garching, Germany}
\affiliation{Munich Center for Quantum Science and Technology (MCQST), 80799 Munich, Germany}

\author{Kritsana Srakaew}
\affiliation{Max-Planck-Institut f\"{u}r Quantenoptik, 85748 Garching, Germany}
\affiliation{Munich Center for Quantum Science and Technology (MCQST), 80799 Munich, Germany}

\author{David Wei}
\affiliation{Max-Planck-Institut f\"{u}r Quantenoptik, 85748 Garching, Germany}
\affiliation{Munich Center for Quantum Science and Technology (MCQST), 80799 Munich, Germany}

\author{Johannes Zeiher}
\affiliation{Max-Planck-Institut f\"{u}r Quantenoptik, 85748 Garching, Germany}
\thanks{present address: Department of Physics, University of California, Berkeley, California 94720, USA}
\affiliation{Munich Center for Quantum Science and Technology (MCQST), 80799 Munich, Germany}

\author{Christian Gross}%
\affiliation{Max-Planck-Institut f\"{u}r Quantenoptik, 85748 Garching, Germany}
\affiliation{Munich Center for Quantum Science and Technology (MCQST), 80799 Munich, Germany}
\affiliation{Physikalisches Institut, Eberhard Karls Universit\"{a}t T\"{u}bingen, 72076 T\"{u}bingen, Germany}

\author{Immanuel Bloch}%
\affiliation{Max-Planck-Institut f\"{u}r Quantenoptik, 85748 Garching, Germany}
\affiliation{Munich Center for Quantum Science and Technology (MCQST), 80799 Munich, Germany}
\affiliation{Fakult\"{a}t f\"{u}r Physik, Ludwig-Maximilians-Universit\"{a}t M\"{u}nchen, 80799 M\"{u}nchen, Germany}%

\date{\today}


\begin{abstract}
Precise control and study of molecules is challenging due to the variety of internal degrees of freedom and local coordinates that are typically not controlled in an experiment.
Employing quantum gas microscopy to position and resolve the atoms in Rydberg macrodimer states solves almost all of these challenges and enables unique access to the molecular frame. 
Here, we demonstrate the power of this approach and present first photoassociation studies for different molecular symmetries in which the molecular orientation relative to an applied magnetic field, the polarization of the excitation light and the initial atomic state are fully controlled. 
The observed characteristic dependencies allow for an electronic structure tomography of the molecular state. 
We additionally observe an orientation-dependent Zeeman shift and reveal a significant influence on it caused by the hyperfine interaction of the macrodimer state. 
Finally, we demonstrate controlled engineering of the electrostatic binding potential by opening a gap in the energetic vicinity of two crossing pair potentials.

\end{abstract}

\maketitle

\section{Introduction}
Understanding the interplay of electronic and nuclear dynamics in chemical reactions is an important goal in quantum chemistry and material sciences.
The quantum control available at ultracold temperatures~\cite{Julliene_Review06,bohn_cold_2017,anderegg_optical_2019} has fostered the understanding of this interplay in a series of experiments, ranging from the coherent preparation of rovibrational states~\cite{danzl_quantum_2008,RevModPhys.Feshbach} and the controlled photodissociation into continuum states~\cite{Zelevinsky16} up to the study of ultracold chemical reactions~\cite{ospelkaus_quantum-state_2010,hu_direct_2019,liu_building_2018}.
In such experiments, the quantum uncertainty of the rotational state forbids a fixed orientation of the molecular axis in the laboratory frame~\cite{lefebvre-brion_chapter_2004}.   
Such a molecular alignment can be achieved with strong fields~\cite{RevMod_molecule_alignment,de_miranda_controlling_2011,Denschlag_alignment_prl2014} and enables access to the molecular frame of reference.
This is a prerequisite for probing the electronic wave function of the molecule since the molecular axis acts as its quantization axis.
Molecular bound states of highly-excited Rydberg atoms feature enormous bond lengths extending into the micrometer regime and provide an alternative route to access the molecular frame.
The large size and accordingly small rotational energy splittings of these so-called Rydberg macrodimers~\cite{Boisseau2002,Sassmannshausen2016,Overstreet2009} enables the excitation of molecules from atom pairs aligned in the laboratory frame and held at a distance matching the bond length by individual optical traps\,\cite{Macrodimers_singleatoms_2019}. 
Furthermore, the small binding energies allow to engineer and shape the molecular potentials already by weak external forces and fields~\cite{Casimir_Macrodimers_Rostock}.
\begin{figure*}[htp]
  \centering
  \includegraphics[width=1.5\columnwidth]{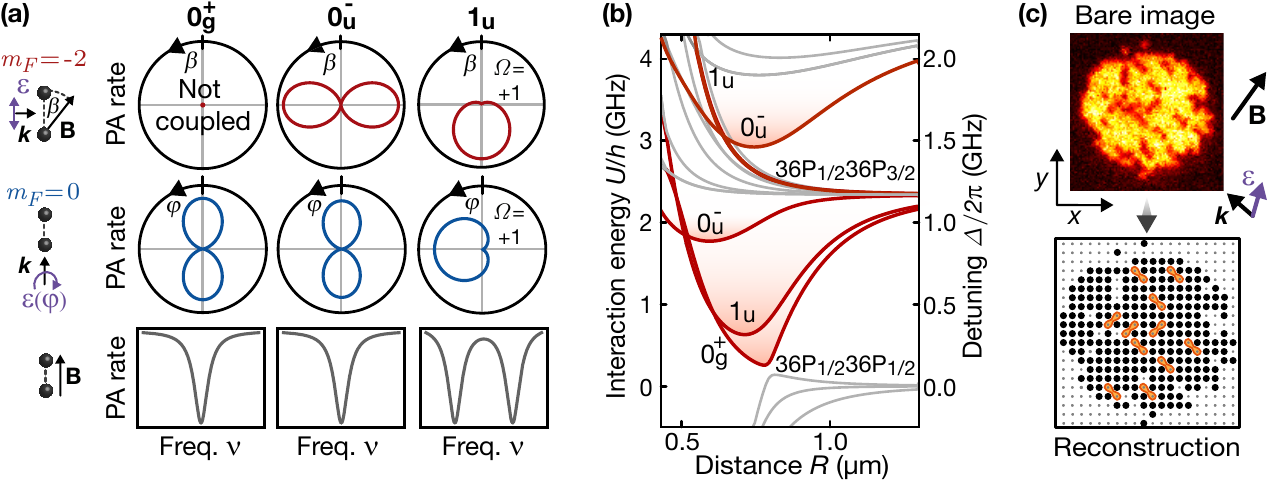}%
  \caption{\label{fig:1}
Overview of the experiment. (a) The photoassociation (PA) rate to molecular potentials $|\Omega|^{\pm}_{g/u}$ depends characteristically on the angle $\beta$ between molecular axis and magnetic field, the initial state $|F=2,m_F\rangle$ and the circular polarization angle $\varphi$ of the polarization $\vec{\varepsilon}$ of the PA laser. For $|\Omega| \neq 0$, the vibrational resonances experience a Zeeman splitting proportional to the magnetic field projection on the molecular axis. (b) Blue-detuned from the atomic $36P_{1/2}$ transition, we find $0^-_u$,$0^+_g$ and $1_u$ molecular potentials. (c) Starting with a near unity filled two-dimensional atom array, we excite pairs of ground state atoms at lattice diagonal distance into molecular macrodimer states which leave the system afterwards. The polarization $\vec{\varepsilon}$ and the magnetic field are tunable. The bond lengths studied here roughly match the diagonal distance of the array. The reconstructed images contain a correlation signal at separation vectors $\mathbf{R} = (1,\pm1)a_{\mathrm{lat}}$ perpendicular\,(parallel) to the $\mathbf{k}$-vector of the PA laser, each corresponding to a certain orientation relative to $\mathbf{B}$ and $\vec{\varepsilon}$.}
\end{figure*}

Macrodimers~\cite{Samboy2011,Samboy2011a,schwettmann_analysis_2007,Shaffer2018} are electrostatically bound states in local minima of Rydberg interaction potentials~\cite{Weber2017,Deiglmayr2016}.
These states are well described by electronic quantum numbers $|\Omega|^{\pm}_{g/u}$, with $\Omega = m^\prime_{J_1}+m^\prime_{J_2}$ the total electronic angular momentum  projection of both bound atoms along the interatomic axis and the superscript\,(subscript) denoting the reflection\,(inversion) symmetry~\cite{Stanojevic2006,Weber2017}. 
Due to the huge polarizability of Rydberg atoms, these molecules feature bond lengths up to the micrometer scale, typically one order of magnitude larger than the extension of the Rydberg wave function itself~\cite{gallagher_1994}. 
The absence of overlapping orbitals classifies them as purely long-range molecules~\cite{Julliene_Review06} and significantly simplifies the calculation of binding potentials and molecular wave functions $|\Psi_\mathrm{Mol};\Omega^\pm_{g/u}\rangle = \sum_{i,j}c_{ij}|r_{i}r_{j}\rangle$ from ab initio principles. 
Knowledge of the decomposition in asymptotic Rydberg pair states $|r_{i}r_{j}\rangle $ enables precise calculations of excitation rates~\cite{Samboy2011} and response to external fields, which both depend on the molecular orientation. 
The excitation of \emph{aligned} pairs of unbound atoms to these huge \emph{aligned} molecules allows one to map out these dependencies, as summarized in Fig.~\ref{fig:1}\,(a).

Here, we combined such spatially resolved photoassociation~(PA) studies in the molecular frame with precision spectroscopy in order to benchmark and identify different molecular symmetries~\cite{pure_long_range_1,pure_long_range_2}.
The observed PA rates depend on the angle $\beta$ between the initial quantization axis of the atoms, given by the magnetic field $\textbf{B}$, and the reference frame of the associated molecule, defined by the interatomic axis $\textbf{R}$.
Resolving these dependencies realizes an electronic structure tomography of the molecule as they expose the underlying molecular quantum numbers.
Additionally, we observed a first order Zeeman interaction of the interaction potentials with a magnetic field aligned with $\mathbf{R}$, which is absent for a purely transverse field.
The corresponding energy shifts of the vibrational resonances are another fingerprint of the electronic structure and enable the selective molecular alignment by the frequency of the PA light.  
Finally, using the transverse field, we shaped the binding potentials by breaking the symmetry of the molecule and coupling two crossing pair potentials with a tunable coupling strength, which modifies the binding potential and leads to controlled predissociation~\cite{Predissociation_Chemical_Rev}.
\section{Experimental setup}
Our experiments started with a two-dimensional atomic Mott insulator of $^{87}\mathrm{Rb}$ atoms loaded into a square optical lattice, with lattice constant $a_{\mathrm{lat}}=532\,$nm and near unity filling of $94(1)\%$~\cite{Sherson2010}. 
The molecular potentials studied here are shifted by the interaction energy $U$ relative to the asymptotic state $|36P_{1/2}36P_{1/2}\rangle$, see Fig.~\ref{fig:1}\,(b).
The two-photon PA of pairs of ground state atoms $|gg\rangle$ (with $|g\rangle$ the specific ground state of the sample) to molecular states $|\Psi_\mathrm{Mol};\Omega^{\pm}_{g/u}\rangle$ leads to resonances at laser detunings $\Delta/2\pi = U/2 h$ relative to the $36P_{1/2}$ transition. The excitation was driven by an ultraviolet (UV) laser at $\lambda = 298\,$nm sent along the diagonal of the lattice~\cite{Macrodimers_singleatoms_2019}. 
The molecular bond length close to the diagonal distance of the lattice and the strong confinement provides significant Franck-Condon overlap at a distance of $\sqrt{2}a_{\mathrm{lat}}$.
The spatial ordering of the ground state atoms on the lattice ensures the alignment of the associated molecules along the two possible diagonal directions. A subsequent dephasing of the contributing rotational modes can be neglected due to the rotational constant~$B_{\mathrm{rot}} \approx 200\,$Hz~\cite{Demtroeder_rotation_2007}, which is smaller than radiative decay rate of the macrodimer state.
The UV polarization, the magnetic field amplitude and its direction relative to the lattice remain tuning parameters, see Fig.~\ref{fig:1}\,(c).
Because the Rydberg atoms are efficiently ejected from the optical lattice~\cite{Macrodimers_singleatoms_2019,Bernien2017}, molecular excitation leads to correlated atom loss at a distance of $\sqrt{2}a_{\mathrm{lat}}$, which can be revealed microscopically by imaging the remaining atoms with our high-resolution objective~\cite{Sherson2010,Bakr2009}.
We quantified the photoassociation rate by evaluating ensemble-averaged hole-hole correlations $g^{(2)}(\delta x,\delta y) = (\langle \hat{h}_{k+\delta x,l+\delta y}\hat{h}_{k,l} \rangle - \langle \hat{h}_{k+\delta x,l+\delta y}\rangle\langle\hat{h}_{k,l} \rangle)_{k,l}$. Here, $\hat{h}_{k,l} = 1 - \hat{n}_{k,l}$ is the hole operator at site $(k,l)$, $\hat{n}_{k,l}$ is the atom number operator, which is 1\,(0) for an occupied\,(empty) site, and $(\,)_{k,l}$  denotes averaging over all sites.

\begin{figure*}
\centering
\includegraphics[width=1.8\columnwidth]{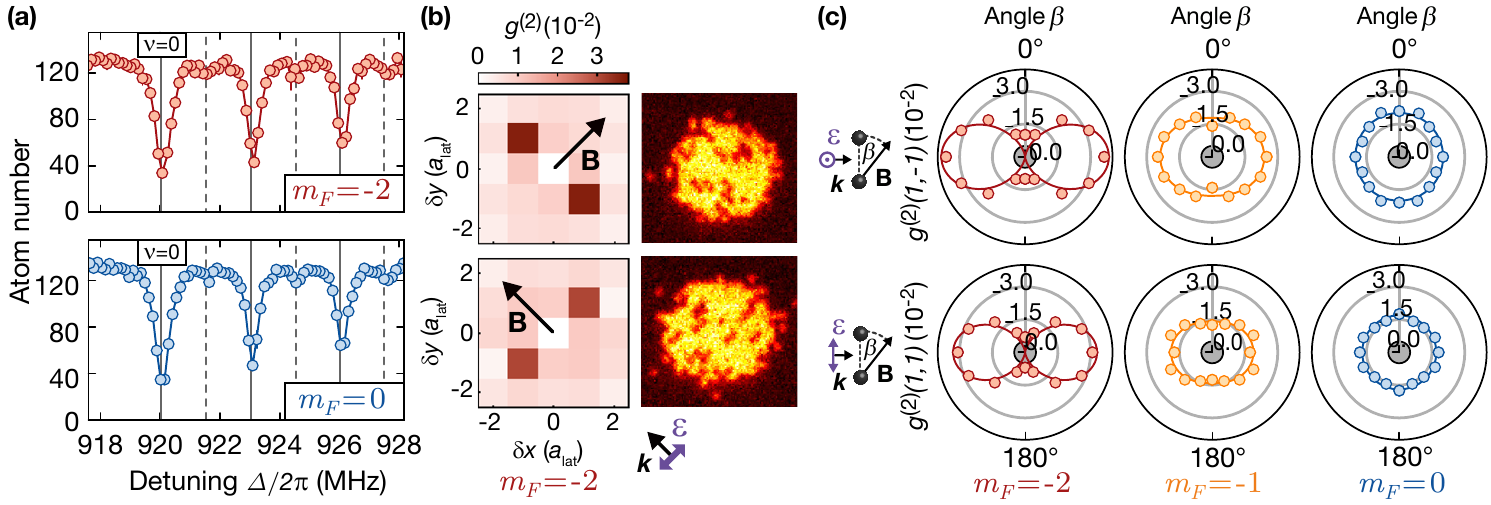}%
  \caption{ \label{fig:2}
  Microscopic excitation signatures for $0^-_u$ molecules. (a) The spectroscopic photoassociation signal can be observed starting from $|m_F=-2\rangle$ (red) and $|m_F=0\rangle$ (blue) and matches the calculated energies for even\,(odd) vibrational quantum numbers $\nu$, indicated by solid\,(dashed) gray lines. (b) Comparing the hole-hole correlations $g^{(2)}(\delta x, \delta y)$ starting from $|m_F=-2\rangle$ for two orthogonal magnetic field orientations, we observe that excitation only occurs for atom pairs oriented perpendicular to the magnetic field. For both cases, we show two exemplary images from the quantum gas microscope from which the correlation strengths are derived. (c) Rotating the orientation of the magnetic field, we find excitation curves characteristic for $0^-_u$, which depend on the initial state and the polarization $\vec{\varepsilon}$. As shown in (b), $\vec{\varepsilon}$ was parallel\,(perpendicular) to $\mathbf{R}$ for $g^{(2)}(1,\pm 1)$. Solid lines are theoretically expected angular dependencies, where the overall amplitude was left as a fitting parameter. All error bars on the data points denote one standard error of the mean (s.e.m.) and grey circles indicate an expected background correlation signal.}
\end{figure*}
\section{Competing reference frames}
In a first experiment, we characterized the $0^-_u$ potential shifted by $1.84\,$GHz relative to the reference asymptote, see Fig.~\ref{fig:1}\,(b).
The spectroscopic signal in the total atom number is similar for all ground states within the $F=2$ hyperfine manifold, as explicitly shown for $m_F=-2$ and $m_F=0$ in Fig.~\ref{fig:2}\,(a).
As expected from Franck-Condon factors, we generally observe that PA rates are maximal for the lowest vibrational quantum number $\nu$ and are higher for even compared to odd $\nu$\,\cite{Macrodimers_singleatoms_2019}.
Our microscopic resolution, however, is expected to reveal differences in the alignment of the associated molecules, also dependent on the magnetic field orientation.
To show this, we first prepared ground state atoms in $\ket{m_F=-2} = \ket{m_J=-1/2} \otimes \ket{m_I=-3/2}$, with $m_J$ and $m_I$ the electronic and nuclear spin projection on the magnetic field. 
Then, we illuminated the atoms with UV light, linearly polarized in the atomic plane and on resonance with the lowest vibrational macrodimer state, until about three molecules were excited.
By adding a finite magnetic field of $B=1\,$G aligned along either of the lattice diagonals, we observe that photoassociation only occurs perpendicular to the magnetic field, as shown in Fig.~\ref{fig:2}\,(b). 
This is a direct consequence of the interplay between the different quantization axes of the ground state atoms and the molecule. 
Due to the small hyperfine interaction of Rydberg atoms, only the summed fine-structure state $|M_J\rangle$ in the ground state $\ket{m_F=-2} \otimes \ket{m_F=-2} =\ket{M_J=-1} \otimes \ket{M_I= -3}$ takes part in the coupling, while the nuclear part $|M_I\rangle$ does not contribute. 
Characterizing the excitation in the molecular frame, the UV light is $\pi$-polarized for $\vec{\varepsilon}\parallel\mathbf{R}$ (i.e. $g^{(2)}(1,1)$) and a superposition of $\sigma^\pm$-components for $\vec{\varepsilon}\perp\mathbf{R}$ (i.e. $g^{(2)}(1,-1)$). 
In both cases, dipole selection rules do not allow a transition from $M_J=-1$ to $\Omega=0$ by the absorption of two photons. 
However, for a finite angle $\beta$ between $\mathbf{R}$ and $\mathbf{B}$, the initial state has to be rotated into the molecular frame, which changes its electronic decomposition and enables molecular excitation.

In a more detailed study, we varied the relative angle $\beta$ between the atom pairs and the magnetic field in five steps from $0^\circ$ to $90^\circ$. We then quantified the PA signal by the observed hole-hole correlations along both diagonals for different initial states, as shown in Fig.~\ref{fig:2}\,(c).
In agreement with our findings of the previous paragraph and Fig.~\ref{fig:1}\,(a), PA rates starting from $|m_F=-2\rangle$ vanish for $\beta = 0$ and are maximal for $\beta = 90^\circ$. 
Starting from $|m_F=-1\rangle$, we find a strikingly different $\beta$-dependence, again consistent with the calculation. 
For $|m_F=0\rangle$, the photoassociation does not depend on $\beta$ any longer.
The observed angular distributions are characteristic for $0^-_u$ potentials and can be attributed to the angular dependent electronic decomposition of the ground state in the molecular frame, see Appendix~\Aref{A}. 
They provide additional information not available in the simple spectroscopic data and allow to identify the symmetry of the underlying molecular potential. 
Furthermore, we found that the PA rates in Fig.\,\ref{fig:2}\,(c) for $\vec{\varepsilon}\parallel\mathbf{R}$ reach only $83\%$ of the value for $\vec{\varepsilon}\perp\mathbf{R}$. 
This is close to the theoretical value of $87\%$ for this specific $0^-_u$ potential, see table~\ref{table:Table:3}. 
A similar measurement for a previously studied $0^+_g$ potential~\cite{Macrodimers_singleatoms_2019} reveals that the excitation in this case is not possible from $|m_F=-2\rangle $, independent of the angle $\beta$~(see Appendix~\Aref{A}).
\begin{figure}
\centering
\includegraphics[width=1.0\columnwidth]{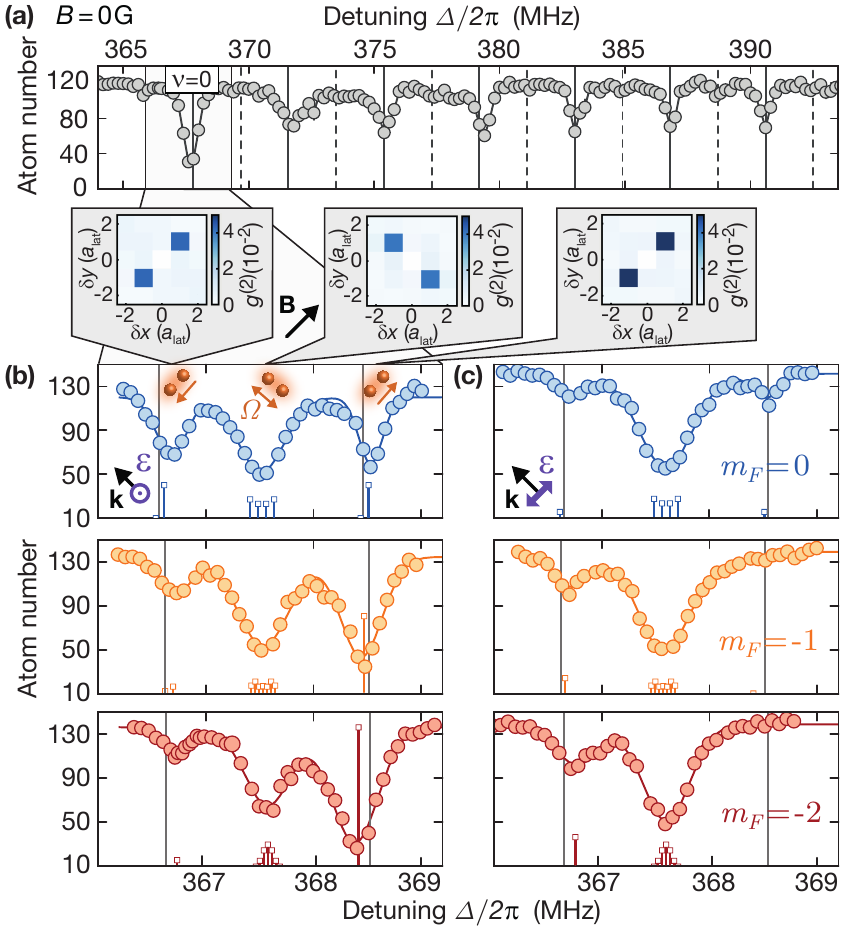}%
  \caption{ \label{fig:3}
Excitation signatures for $1_u$ molecules. (a) At zero magnetic field, we find a series of narrow vibrational resonances matching the theoretical energies\,(grey). (b,c) Measurements for a finite magnetic field and a light field oscillating perpendicular and parallel to the atomic plane reveal a splitting of the lines. The outer resonances correspond to a molecular alignment $ \mathbf{R}\parallel\mathbf{B}$ and the central one to $ \mathbf{R}\perp\mathbf{B}$, as shown in the recorded correlation signals. 
The difference between the data and a calculation only based on the electronic state decomposition\,(grey solid line) can be explained by the hyperfine interaction\,(colored bars).
The height of the bars indicate the calculated relative excitation rates, which are also in agreement with the observations. All error bars on the data points denote one s.e.m.}
\end{figure}
\section{Zeeman and hyperfine interaction}
In addition to the strong orientation-dependent coupling rates, molecular states with $|\Omega| \neq 0$ are expected to also energetically split in the presence of a magnetic field.  
At zero field, the calculated resonance positions for the $1_u$ potential located $735$\,MHz above the reference asymptote (see Fig.~\ref{fig:1}\,(b)) agree with the measured results, see Fig.~\ref{fig:3}\,(a).
In this configuration, both possible molecular orientations $|\Psi_\mathrm{Mol};\pm 1_u\rangle$ are degenerate.
Measurements at a finite magnetic field $B = 2.0\,$G applied along the lattice diagonal perpendicular to the UV propagation direction, with initial states $m_F=-2,-1,0$ and an excitation light field oscillating perpendicular and parallel to the atomic plane, are shown in Figs.~\ref{fig:3}\,(b) and~\ref{fig:3}\,(c).
In contrast to Fig~\ref{fig:1}\,(a), the two possible orientations of the molecules lead to a splitting of the vibrational resonances into three instead of two lines. 
While the two outer lines $|\Psi_\mathrm{Mol};\pm 1_u\rangle_{\parallel}$ correspond to $\beta = 0$, molecules $|\Psi_\mathrm{Mol};\pm 1_u\rangle_{\perp}$ created at the central unshifted resonance are aligned perpendicular to $\mathbf{B}$ with $\beta = 90^\circ$.
This is supported by calculations including the interaction of the molecular state with the magnetic field, which predicts a first order Zeeman shift only for the molecules aligned with the magnetic field.
The observed PA rates for different polarizations and initial states can again be predicted by the molecular state decomposition and the contributing Clebsch-Gordan coefficients.
As expected, we find that both states $|\Psi_\mathrm{Mol};\pm 1_u\rangle_{\parallel}$ are generally coupled more strongly for $\sigma^\pm$-polarization compared to $\pi$-polarization (compare Figs.~\ref{fig:3}\,(b) and (c)).
In agreement with Fig.~\ref{fig:1}\,(a), the state $|\Psi_\mathrm{Mol};+ 1_u\rangle_{\parallel}$ in Fig.~\ref{fig:3}\,(c) with $\Omega =+1$ cannot be coupled with $\pi$-polarized light from $|m_F = -2\rangle$ atoms because the dipole matrix elements starting from $|M_J = -1\rangle$ vanish.
\\
Interestingly, the calculated splitting shows a small but significant deviation from the measurements, which depends on the initial state and the UV polarization.
For $|m_F=0\rangle$ as a starting state, we observe an asymmetric splitting, while the splitting for $|m_F=-2\rangle$ is symmetric but overestimated by the theory. 
Both signatures are even more pronounced for a $1_u$ potential at lower principal quantum numbers, see Appendix~\Bref{B}.
Extending the theory to include the hyperfine interaction of the contributing asymptotic Rydberg pair states $|r_i r_j\rangle$, we are able to also account for the remaining deviation. 
To our knowledge, this is the first observation of hyperfine interactions in Rydberg interaction potentials, enabled by our high spectroscopic resolution and the narrow vibrational resonances of Rydberg macrodimers.
The additional observation that the central $|\Psi_\mathrm{Mol};\pm 1_u \rangle_{\perp}$ resonance is broader than the Zeeman split resonances is a consequence of the two competing reference frames.
Rotating the initial states by $\beta = 90^\circ$ into the molecular frame, we find a large contribution of different nuclear spin orientations, which are all split by the hyperfine interaction.

\section{Potential shaping and predissociation}
\begin{figure}[htb]
  \centering
  \includegraphics[width=0.5\textwidth]{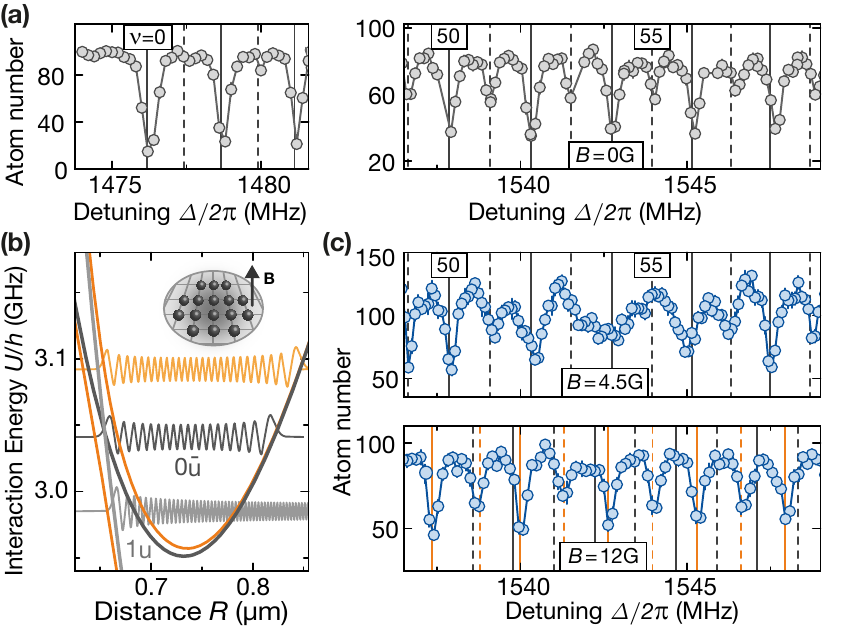}%
  \caption{\label{fig:4}
  Potential engineering and controlled predissociation. (a) At zero field, the calculated eigenmodes~(grey lines) within the $0^-_u$ potential match the observations. (b) A magnetic field perpendicular to the molecular axis couples the $0^-_u$ potential (dark grey) to a crossing $1_u$ potential (light grey), forming a new combined binding potential (here shown for $B=25\,$G in orange). (c) At $4.5\,$G, the vibrational resonances at certain frequencies above the crossing are significantly broadened. At $12\,$G, we again find a clean spectrum which coincides with the newly engineered potential. Grey\,(orange) lines represent the calculated eigenmodes in the isolated $0^-_u$ (combined) potential. Due to an overall energy shift, both sets of eigenmodes were manually overlapped with the lowest line of the spectrum. All error bars on the data points denote one s.e.m.}
\end{figure}

In a final experiment, we show how the small binding energies of macrodimers enable control over molecular binding potentials and the internuclear wavefunction.
For this purpose, we focus on a $0^-_u$ potential in the vicinity of a crossing repulsive $1_u$ potential, shown in the upper part of Fig.~\ref{fig:1}\,(b). A spectroscopy of the vibrational modes in the $0^-_u$ potential at zero field is presented in Fig.~\ref{fig:4}\,(a).
A magnetic field $\mathbf{B} \parallel \mathbf{R}$ shifts both states within the $1_u$ potential but the system still obeys the rotational symmetry of the molecule, leaving both potentials uncoupled. 
This changes for a finite transverse magnetic field $\mathbf{B} \perp \mathbf{R}$ where the symmetry is broken and $\Omega$ is not conserved anymore. 
Now, a tunable Zeeman coupling between both potentials emerges, which is proportional to the magnetic field amplitude and expected to affect the vibrational motion, see Fig.~\ref{fig:4}\,(b).
To study this effect, we initialize our atoms in $m_F=0$ and measure the vibrational series again for a finite magnetic field orthogonal to the atomic plane.
For $B=4.5$\,G, we now observe that some of the macrodimer lines \glqq blur out\grqq ~in a certain frequency range above the crossover, see Fig.~\ref{fig:4}\,(c).
We attribute this to the non-adiabatic coupling of the bound $0^-_u$ vibrational modes to the continuum modes of the repulsive $1_u$ potential, which significantly decreases their lifetime and broadens their spectral lines.
This phenomenon is called predissociation~\cite{Predissociation_Chemical_Rev,lefebvre-brion_chapter_2004} and its tunability by external magnetic fields has been studied~\cite{kato_line_1993,vigue_j_natural_1981}.
Compared to previous studies, our much lower vibrational energies and the well defined angle $\beta$ in our system allow to study this at lower field amplitude and with unprecedented dynamic range and controllability.
At $B=12$\,G, we recover a clean vibrational spectrum with well defined resonances, however with a significant increase of the experimentally observed vibrational spacing from $\Delta\nu_{\mathrm{exp}} = 2.45(1)\,$MHz to $\Delta\nu_{\mathrm{exp}} = 2.66(1)\,$MHz, see Fig.~\ref{fig:4}\,(c). 
This is in quantitative agreement with the calculated values $\Delta\nu_{0^-_u}=2.45(1)\,$MHz and $\Delta\nu_{\mathrm{comb}}=2.66(2)\,$MHz in the isolated $0^-_u$ potential and the new combined potential, indicating that the interatomic motion now follows the avoided crossing adiabatically.
Additionally, we observe an overall energy offset, which also quantifies the influence of the magnetic field on the binding potential.
Further data in the regime between $B=3.5$\,G and $B=6.5$\,G, where a breakdown of the Born-Oppenheimer approximation leads to predissociation, is shown in Appendix~\Cref{C}.
\section{Conclusion}
Our observations show how the study of Rydberg macrodimers in quantum gas microscopes enable characterization and identification of molecular symmetries by their microscopic couplings at a level of control not present in conventional molecule platforms. 
Within the spectrum of Zeeman-split $1_u$ molecular transitions, we observe a hyperfine interaction between the macrodimer state and both nuclei.
Furthermore, we show how the binding potentials can be modified with external magnetic fields. 
In future studies, potential shaping might also be extended by coupling neigbhoring Rydberg states with microwave fields~\cite{MW_control_Petrosyan,sevincli_microwave_2014}.
For quantum simulation purposes, the strongly directional coupling rates to molecular states can be used to engineer anisotropic interaction potentials, also in the context of Rydberg dressing~\cite{Bijnen2015,Jau2016,Zeiher2016a,Coherent_Zeiher,Dressing_Schleyer-Smith}. 
In particular, the tunability with the light polarization enables very fast switching of admixed interactions.
Finally, by choosing magnetic field and light polarization such that coupling rates reach a maximum, one might observe novel four-body interactions~\cite{four_body_FRET} arising between all four Rydberg atoms contributing to pairs of macrodimers.

\begin{acknowledgements}
\textbf{Acknowledgements:}
  We thank all contributors to the open-source programs ``pair interaction'' and ``ARC'' as well as Sebastian Weber, Dan M. Stamper-Kurn, Valentin Walther, Simon Evered, Andreas Kruckenhauser and Mathieu Barbier for valuable discussions.
We acknowledge funding by the Max Planck Society (MPG) and from Deutsche Forschungsgemeinschaft (DFG, German Research Foundation) under Germany’s Excellence Strategy – EXC-2111 – 390814868 and Project No. BL 574/15-1 within SPP 1929 (GiRyd). 
This project has received funding from the European Union’s Horizon 2020 research and innovation programme under grant agreement No. 817482 (PASQuanS) and the European Research Council (ERC) No. 678580 (RyD-QMB).
K.S. acknowledges funding through a stipend from the International Max Planck Research School (IMPRS) for Quantum Science and Technology and J.R. acknowledges funding from the Max Planck Harvard Research Center for Quantum Optics. 
\end{acknowledgements}

\section*{APPENDIX A: \textbf{Photoassociation}}\label{A}
\subsection*{1. \textbf{Rotation into the molecular frame}}
\begin{figure*}
\centering
\includegraphics[width=0.75\textwidth]{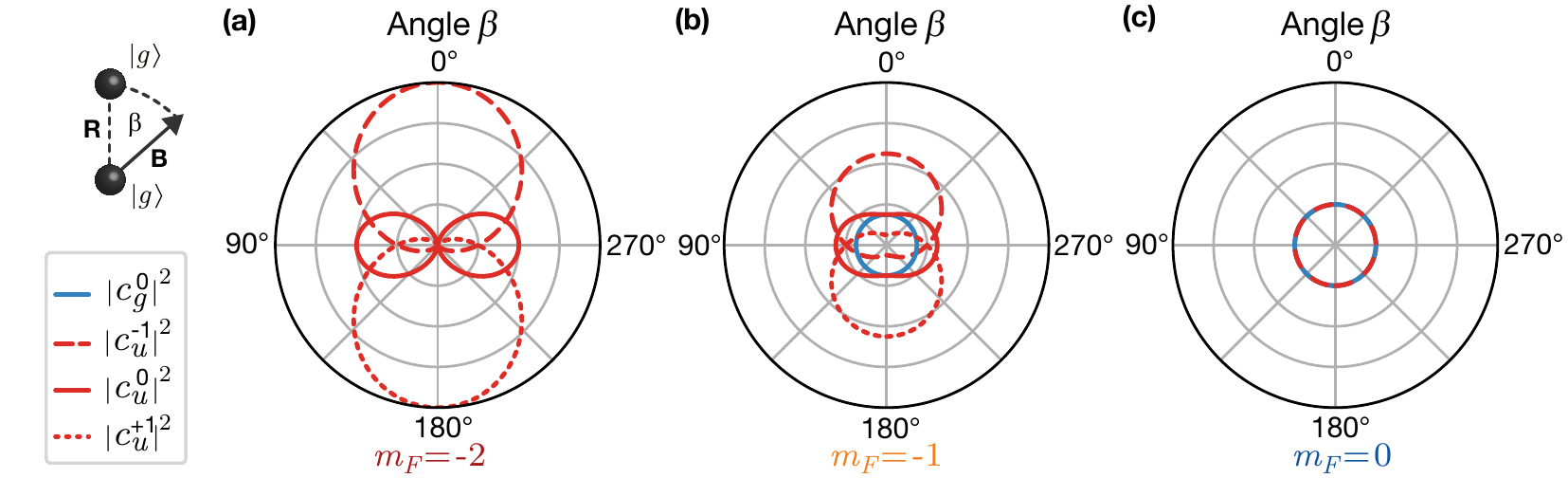}%
  \caption{ \label{supfig:1}
Initial state decomposition in inversion eigenstates.
(a) For $|g\rangle = |2,-2\rangle$ and $\beta = 0^\circ\,(180^\circ)$, only $|-1_u\rangle_J\,(|+1_u\rangle_J)$ contributes. For $\beta = 90^\circ$, the initial pair state is a superposition of all three ungerade states. There is no contribution from $|0_g\rangle_J$. (b) For $|g\rangle = |2,-1\rangle$, the tendency is similar but the dependence turns out to be weaker. Additionally, $c_g^0$ is finite but independent of $\beta$. (c) For $|g\rangle = |2,0\rangle$, all states contribute with the same amplitude for all angles.}
\end{figure*}
Rabi frequencies between two internal atomic states are usually calculated in the reference frame of the atom, here given by the magnetic field $\mathbf{B}$. 
For molecules, the interatomic axis $\mathbf{R}$ enters as an additional parameter. 
We account for this by transforming the $^{87}\mathrm{Rb}$~\cite{Rb87_Steck} ground state $|g\rangle_\mathbf{B} = |F,m_F\rangle$ into the frame of the molecule by a rotation operator $\hat{U}(\beta)$, 
\begin{equation}\renewcommand\theequation{A\arabic{equation}}\label{eq:rotation_init}
|g\rangle_{\mathbf{B}} \rightarrow \hat{U}(\beta)|g\rangle_{\mathbf{B}} = \sum_{m_F} c^{F}_{m_F}|F,m_F\rangle \equiv |g\rangle_{\mathbf{R}}.
\end{equation} 
Here, $\beta$ is the angle between $\mathbf{B}$ and $\mathbf{R}$.
Accordingly, the light polarization is expressed in the molecular frame.
Alternatively, the molecular state $| \Psi_{\mathrm{Mol}};\Omega^{\pm}_{g/u}\rangle_{\mathbf{R}} \rightarrow \hat{U}(-\beta)|\Psi_{\mathrm{Mol}};\Omega^{\pm}_{g/u}\rangle_{\mathbf{R}} \equiv |\Psi_{\mathrm{Mol}};\Omega^{\pm}_{g/u}\rangle_{\mathbf{B}} $~\cite{Samboy2011,Samboy2011a,Macrodimers_singleatoms_2019} can be rotated into the atomic frame. 
Because the molecular state consists of a large number of atomic pair states which have to be rotated individually, we rotate the initial state using Eq.~\ref{eq:rotation_init}.

\subsection*{2. \textbf{Decomposition of the initial state}}
\begin{figure}
\centering
\includegraphics[width=1\columnwidth]{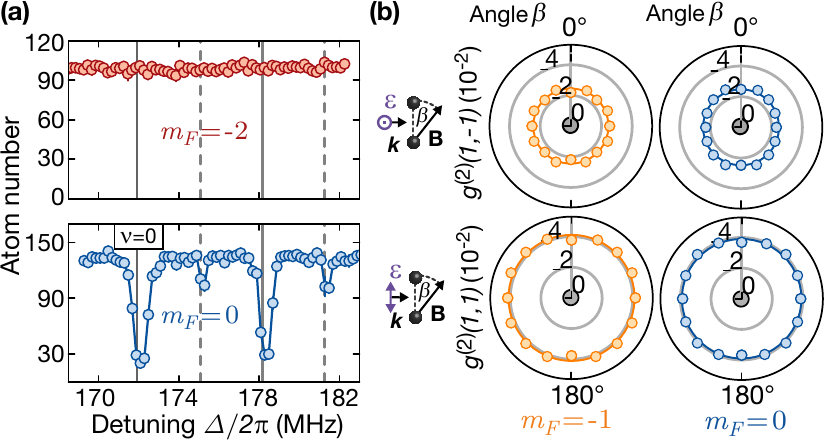}%
  \caption{ \label{supfig:2}
Experimental signatures for $0^+_g$ molecules (a) As expected, excitation is not possible starting from $m_F=-2$. (b) Also consistent with our calculations, the measured correlation strengths are independent of $\beta$ and are stronger for $\vec{\varepsilon}\parallel\mathbf{R}$. Again, the angular dependence was adopted from a theoretical calculation (see Fig.~\ref{supfig:1}\,(b,c)) and only the overall signal strength was left as a fit parameter.  All error bars on the data points denote one standard error of the mean (s.e.m.) and grey circles indicate an expected background correlation signal.}
\end{figure}

Due to the small hyperfine interaction strength of Rydberg states, the electronic state of Rydberg macrodimers is expressed in the fine structure basis. Decomposing the rotated initial pair state $|gg\rangle_{\mathbf{R}}$ in this basis yields
\begin{equation*}\renewcommand\theequation{A\arabic{equation}}
\begin{split}
 |gg\rangle_{\mathbf{R}} =  & \sum_{\substack{m_{J_1},m_{J_2} \\ \in \left\lbrace\uparrow,\downarrow \right\rbrace}}|m_{J_1},m_{J_2}\rangle \sum_{\substack{m_{I_1},m_{I_2}\in \\ \left[-3/2,3/2 \right]}}\mathcal{C}^{1/2\,\,\,\,F\,\,\,\,\,3/2}_{m_{J_1}\,m_{F_1}\,m_{I_1}}\\ & \mathcal{C}^{1/2\,\,\,\,F\,\,\,\,3/2}_{m_{J_2}\,m_{F_2}\,m_{I_2}}c^{F}_{m_{F_1}}c^{F}_{m_{F_2}} |m_{I_1},m_{I_2}\rangle,
\end{split}
\end{equation*}
with $\uparrow(\downarrow)$ = +1/2\,(-1/2) for $m_{J_{1}}$ and $m_{J_{2}}$, reordered electron and nuclear angular momenta and Clebsch-Gordan coefficients $\mathcal{C}^{1/2\,F\,3/2}_{m_J\,m_F\,m_I}=\langle 1/2,m_J,3/2,m_I |F,m_F \rangle $. 
Furthermore, we set $m_{F_{1(2)}}=m_{J_{1(2)}}+m_{I_{1(2)}}$ and omit the sum because the Clebsch-Gordan coefficients vanish otherwise. 
After introducing coefficients 
\begin{equation*}\renewcommand\theequation{A\arabic{equation}}
c_{m_{J_1},m_{J_2}}^{\,2} = \sum_{\substack{m_{I_1}\\ m_{I_2}}}\left(
C^{1/2\,\,\,\,F\,\,\,\,\,3/2}_{m_{J_1}\,m_{F_1}\,m_{I_1}}
C^{1/2\,\,\,\,F\,\,\,\,3/2}_{m_{J_2}\,m_{F_2}\,m_{I_2}}
c^{F}_{m_{F_{1}}}c^{F}_{m_{F_{2}}}\right)^2
\end{equation*}
and normalizing nuclear spin states $|\Psi^{m_{J_1},m_{J_2}}\rangle_{I}$, this can be simplified to
\begin{equation*}\renewcommand\theequation{A\arabic{equation}}
\begin{split}
|gg\rangle_{\mathbf{R}} & =  c_{\downarrow\downarrow}\ket{\downarrow\downarrow}_J |\Psi^{\downarrow\downarrow}\rangle_{I} +  c_{\uparrow\uparrow}\ket{\uparrow\uparrow}_J |\Psi^{\uparrow\uparrow}\rangle_{I} \\ & + c_{\downarrow\uparrow}\left(\ket{\downarrow\uparrow}_J |\Psi^{\downarrow\uparrow}\rangle_{I}  + \ket{\uparrow\downarrow}_J |\Psi^{\uparrow\downarrow}\rangle_{I}\right) ,
\end{split}
\end{equation*}
with $c_{\uparrow\downarrow} = c_{\downarrow\uparrow}$ for both atoms populating the same ground state. 
Since dipole-allowed single-photon transitions enforce an inversion symmetry flip, our two-photon photoassociation (PA) conserves the inversion symmetry of the initial state~(see discussion of Eq.~(\ref{eq:Inv_exp})). It is therefore convenient to further decompose $|gg\rangle_{\mathbf{R}}$ into gerade\,$(g)$ and ungerade\,$(u)$ inversion eigenstates $|M_{J\,g/u}\rangle$, with $M_J = m_{J_1}+m_{J_2}$ the summed angular momentum projection on $\mathbf{R}$. For ground state $^{87}\mathrm{Rb}$ with orbital angular momentum $L=0$, this leads to pair states  $|0_{g}\rangle = 1/\sqrt{2}\left(\ket{\uparrow\downarrow} - \ket{\downarrow\uparrow}\right)$, $|-1_{u}\rangle = \ket{\downarrow\downarrow}$, $|0_{u}\rangle = 1/\sqrt{2}(\ket{\uparrow\downarrow} + \ket{\downarrow\uparrow})$ and  $|+1_{u}\rangle = \ket{\uparrow\uparrow}$, which are formally equivalent to the singlet and triplet basis states of two coupled spin-1/2 systems. Introducing new normalized nuclear spin states yields
\begin{equation}\renewcommand\theequation{A\arabic{equation}}\label{eq:decomp}
\begin{split}
|gg\rangle_{\mathbf{R}} & =  c^{-1}_{u}\ket{-1_u}_J\otimes|\Psi_{u}^{-1}\rangle_I + c^{0}_{u}\ket{0_{u}}_J\otimes|\Psi_{u}^{0}\rangle_I \\ & + c^{+1}_{u}\ket{+1_u}_J\otimes|\Psi_{u}^{+1}\rangle_I + c^{0}_{g}\ket{0_g}_J\otimes|\Psi_{g}^{0}\rangle_I.
\end{split}
\end{equation}
The $\beta$-dependent coefficients for the three studied ground states are shown in Fig.~\ref{supfig:1}. 
If only one of the four states couples to a molecular potential, the measured angular dependencies of the excitation rates reproduce these curves. 
This is the case for the $0^-_u$ molecules studied in Fig.~\ref{fig:2} where the chosen light polarization only allows excitation from $|0_u\rangle_J$. 
A comparable study for $0^+_g$ is shown in Fig.~\ref{supfig:2}, which can only be excited by the rotationally invariant state $|0_g\rangle_J$. 
Since this component vanishes for $m_F=-2$, PA is only possible from $m_F=-1$ and $m_F=0$. In contrast to Fig.~\ref{fig:2}, we now observe larger PA rates for $\pi$-polarization compared to $\sigma^\pm$-polarization. 
This is in agreement with a calculation based on the electronic decomposition of the state $|\Psi_{\mathrm{Mol}};0^+_g\rangle$ (see also table~\ref{table:Table:3}) as well as with previous studies on the corresponding $0^+_g$ potential for $n=35$ with very similar coupling characteristics~\cite{Macrodimers_singleatoms_2019}.
If more than one of the four states couple, the PA rates were added. Note also that photoassociation projects the nuclear spin state, which can entangle both nuclei.

\subsection*{3. \textbf{Light polarization}}
\begin{figure}
\centering
\includegraphics[width=1\columnwidth]{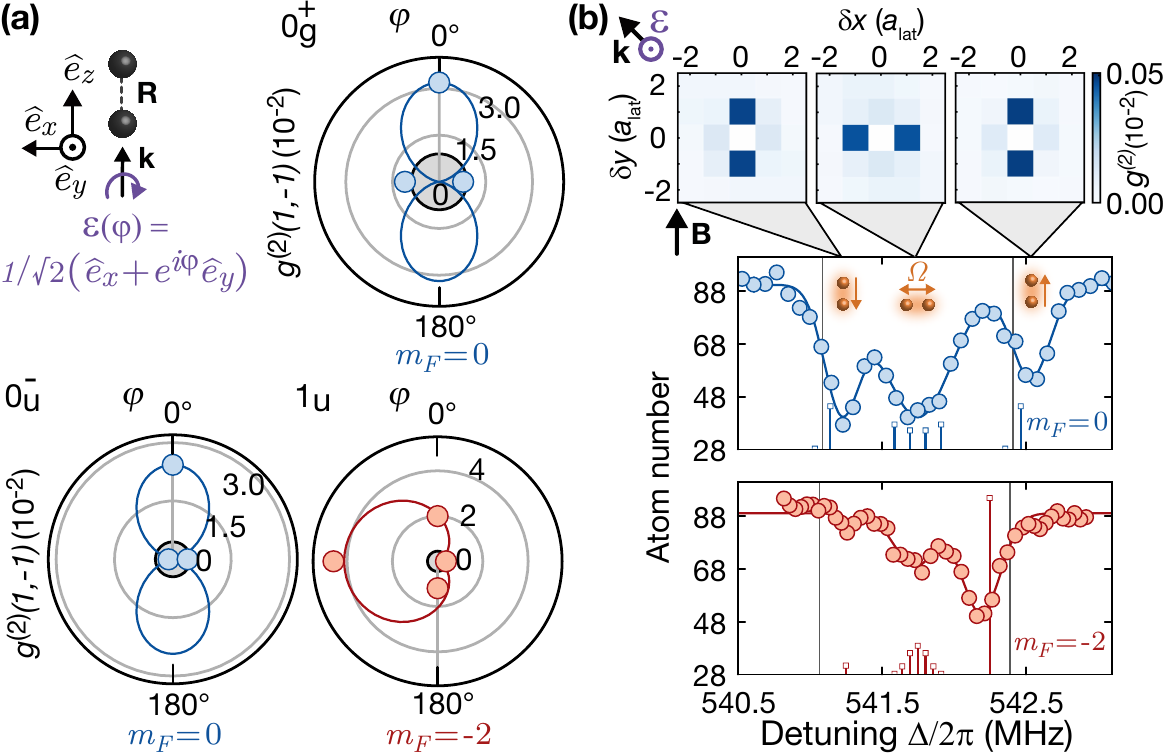}%
  \caption{ \label{supfig:3}
Further details on different molecular symmetries. (a) For $\mathbf{R}\parallel\mathbf{k}$, the three potentials for $n=36$ have characteristic dependencies on the circular polarization angle $\varphi$ of the excitation light.  The angular dependence was adopted by theory and the overall signal amplitude was a fit parameter. Grey areas denote the estimated background correlation signal. (b) For a $1_u$ potential for $n=31$, we again find a splitting of the molecular resonances (here shown for $\nu = 0$) in the presence of a magnetic field of $B=1.5$\,G. Here, $\vec{\varepsilon}$ was directing out of the atomic plane.
Grey lines indicate the expected energies only accounting for the electronic structure of the molecule.
The position\,(height) of the colored bars indicate the resonances\,(scattering rates) after also including the hyperfine interaction.}
\end{figure}

The polarization $\vec{\varepsilon}$ of the excitation light field $\mathbf{E} = E_0\vec{\varepsilon}$ with amplitude $E_0$ is expanded into the spherical basis $\vec{\varepsilon}=\sum_{q\in \{0,\pm1 \} }c_{q}\hat{e}_{q}$ in the molecular frame. We define $\pi$-polarized light $\hat{e}_{0} = \hat{e}_{z}$ for $\vec{\varepsilon}\parallel\mathbf{R}$ and $\sigma^\pm$-polarized light $\hat{e}_{\pm 1} = \mp\frac{1}{\sqrt{2}}\left(\hat{e}_{x} \pm i\hat{e}_{y}\right)$ for a light field oscillating in the plane perpendicular to $\mathbf{R}$.
The PA laser propagates along one lattice diagonal and the molecular bond size studied here is close to the lattice diagonal distance. 
This leads to PA either parallel or perpendicular to the $k$-vector. 
For $\mathbf{R} \perp \mathbf{k}$ and a UV polarization parallel~(perpendicular) to the atomic plane, the polarization vector is $\hat{e}_z$\,($\hat{e}_x$).
For $\mathbf{R} \parallel \mathbf{k}$, the $\pi$-component is always zero and all linear polarizations are equivalent. Now, the phase delay $\varphi$ between $\hat{e}_x$ and $\hat{e}_y$ parametrizes a general polarization vector $\vec{\varepsilon}(\varphi) = \frac{1}{\sqrt{2}}\left(\hat{e}_{x} + e^{i\varphi}\hat{e}_{y}\right)$.  As shown in Fig.~\ref{fig:1}\,(a) and Fig.~\ref{supfig:3}\,(a), this allows us to further characterize the discussed molecular symmetries by the angular momentum conservation in their association. To reveal this, we tuned the UV laser on resonance with the lowest vibrational lines of each potential and compared the correlation strength $g^{(2)}(1,-1)$ which is parallel to the $k$-vector for various $\varphi$. For both potentials with $\Omega = 0$, we find that the PA rates reach a maximum for linear polarization because a combination of $\sigma^+$ and $\sigma^-$ is needed. Coupling the $1_u$ state with $\Omega = +1$ starting from $|m_F=-2\rangle$, the largest coupling can be observed for purely $\sigma^+$-polarized light.
\subsection*{4. \textbf{Calculation of Rabi frequencies}}
In a three-level system, the effective two-photon Rabi frequency $\widetilde\Omega$ coupling an initial state to a final state via an intermediate state reads $\widetilde\Omega = \frac{\widetilde\Omega_1\widetilde\Omega_2}{2\Delta}$. Here, $\widetilde\Omega_{1(2)}$ are the Rabi couplings from the final state to the intermediate state and from the intermediate state to the initial state and $\Delta \gg \widetilde\Omega_{1\,(2)}$ is the intermediate state detuning. The same formalism describes our PA, however, with more than one intermediate state and also several coupled asymptotic pair states within the macrodimer state.
The coupling Hamiltonian reads
\begin{equation*}\renewcommand\theequation{A\arabic{equation}}
\hat{H}_L(\mathbf{E}) = -(\hat{\mathbf{d}}^{(1)}\otimes\mathds{1}^{(2)}+\mathds{1}^{(1)}\otimes \hat{\mathbf{d}}^{(2)})\cdot\mathbf{E},
\end{equation*}
with $\hat{\mathbf{d}}^{\left(1(2)\right)}$ the dipole operators of both individual atoms forming the molecule.
The molecular states $|\Psi_\mathrm{Mol};\Omega^\pm_{g/u}\rangle = \sum_{i,j}c_{ij}(R)|r_{i}r_{j}\rangle$ consist of many asymptotic Rydberg pair states $|r_{i}r_{j}\rangle = |n_i L_i J_i m_{Ji};n_j L_j J_j m_{Jj}\rangle$, mixed by the interatomic interaction. Since our UV laser can only excite Rydberg P-states, only pair states where $L_i = L_j = 1$ contribute.
 \begin{table*}[htp]
  \begin{tabular}{ l  c  c  c  c  c  c  c  c }
    \hline\hline       
    \T\B Dataset \raisebox{9pt}  & state & $\widetilde\Omega_{\mathrm{ref}}/2\pi\,$(MHz) &$\beta$ & $\vec{\varepsilon}$ & $m_F$ & $\gamma_{\mathrm{th}}\,$(Hz) & $t_{\mathrm{uv}}\,$(ms) & $\gamma_{\mathrm{exp}}\,$(Hz) \\ \hline 
    \T\I Fig.~\ref{fig:2}\,(c) \raisebox{9pt}               & $0^-_u$            & 1.25(15)   & 90      & $\hat{e}_0$  & [-2,-1,0]  & \textbf{[1.3(6),\,0.80(39),\,0.64(31)]} & $[20,30,30]$ &  \textbf{[1.7(1),\,0.63(1),\,0.57(1)]} \\     
    \I Fig.~\ref{fig:2}\,(c) \raisebox{9pt}               & $0^-_u$            & 1.25(15)&  90      & $\hat{e}_x$   & [-2,-1,0] & \textbf{[1.5(6),\,0.91(44),\,0.73(36)]} &$[20,30,30]$  & \textbf{[1.9(1),\,0.80(1),\,0.67(1)]} \\    
    \I Fig.~\ref{fig:3}\,(b)\raisebox{9pt}                & $1_u,\Omega=+1$     & 1.50(15) &  0       & $\hat{e}_x$ & 0 & \textbf{23(9)}  &$20$          &  \textbf{2.6(1)} \\     
    \I Fig.~\ref{fig:3}\,(b)\raisebox{9pt}                & $1_u,\Omega=\pm 1$ & 1.50(15) &  90      & $\hat{e}_x$  & 0 & \textbf{45(18)} &$8.0$          &  \textbf{5.1(1)} \\      
    \I Fig.~\ref{fig:3}\,(b) \raisebox{9pt}               & $1_u,\Omega=-1$    & 1.50(15) &  0       & $\hat{e}_x$  & 0 & \textbf{23(9)}  &$20$          &  \textbf{2.2(1)}  \\    
    \I Fig.~\ref{supfig:2}\,(b) \raisebox{9pt}  & $0^+_g$            & 1.15(15) & \ding{56}& $\hat{e}_0$   & [-1,0] & \textbf{[40(21),54(28)]} &$[1.5,0.95]$& \textbf{[26(1),43(1)]} \\ 
    \I Fig.~\ref{supfig:2}\,(b) \raisebox{9pt}  & $0^+_g$            & 1.15(15) & \ding{56}& $\hat{e}_x$   & [-1,0] & \textbf{[13(7),17(9)]}  &$[1.5,0.95]$&  \textbf{[16(1),26(1)]} \\ 
    \I Fig.~\ref{supfig:3}\,(a)                 & $0^-_u$            & 1.50(15) & \ding{56}& $\hat{e}_x$  & 0 & \textbf{1.51(55)}  &$40$          & \textbf{0.60(1)} \\ 
    \I Fig.~\ref{supfig:3}\,(a)                 & $0^+_g$            & 1.50(15) & \ding{56}& $\hat{e}_x$  & 0 & \textbf{49(20)}  &$1.0$         &  \textbf{32(1)} \\ 
    \I\B Fig.~\ref{supfig:3}\,(a)                 & $1_u\,\Omega=+1$   & 1.9(1) &  0       & $\hat{e}_{+1}$& -2 & \textbf{934(197)} &$0.044$      & \textbf{1045(13)} \\     
 \hline\hline  
  \end{tabular}
  \caption{\label{table:Table:3}
Calculated and experimental photoassociation rates. For measurements with varying $\beta$ and $\vec{\varepsilon}$, the value where the hole-hole correlations reach its maximum were included. For linear UV polarization with $\vec{\varepsilon} \perp \mathbf{R}$, we choose the polarization to be along the $x$-direction. For the hyperfine split $1_u$ potentials studied in Fig.~\ref{fig:3}\,(b), the rates of all transitions contributing to the same of the three resolved resonances were added. This involves all hyperfine transitions as well as $\Omega=\pm 1$ for the central resonance. If the coupling was independent of the angle $\beta$, the value is not included. Errors on $\gamma_{\mathrm{th}}$ were estimated by the fluctuations in the reference singe-photon Rabi frequency $\widetilde\Omega_{\mathrm{ref}}$ (see Appendix~\Dref{D}), which contribute to $\gamma_{\mathrm{Mol}}$ to the fourth power. Uncertainties in $\gamma_{\mathrm{exp}}$ originate from errors in the correlation signal $g^{(2)}(\delta x, \delta y)$, obtained with a Bootstrap algorithm (delete-1 Jackknife). Overall, $\gamma_{\mathrm{th}}\,$(Hz) and $\gamma_{\mathrm{exp}}\,$(Hz) are in good agreement. Only for the correlations presented in Fig.~\ref{fig:3}\,(b), the deviation is significant. For this dataset, however, we have indications that the UV beam was slightly misaligned, which strongly decreases the scattering rate.
}
\end{table*}
The states $|r_ir_j\rangle$ are coupled from the ground state $|gg\rangle$ by two-photon transitions via the two intermediate states $|r_ig\rangle$ and $|gr_j\rangle$, yielding a summed contribution
\begin{equation}\renewcommand\theequation{A\arabic{equation}}\label{eq:asympt_pairstate_coupling}
\widetilde\Omega_{ij} = \frac{\widetilde\Omega_{i}\widetilde\Omega_{j}}{2}\left(\frac{1}{\Delta_{i}}+\frac{1}{\Delta_{j}}\right).
\end{equation} 
Here, $\Delta_{i/j}$ are detunings to either $36P_{1/2}$ or $36P_{3/2}$, dependent on $J_{i/j}$, and  $\widetilde\Omega_{i/j} = \frac{1}{\hbar}\langle r_{i/j}|\hat{\mathbf{d}}\cdot \mathbf{E} |g \rangle$ are single particle Rabi frequencies. 
Summing over all contributing states, the total coupling rate $\widetilde\Omega_{\mathrm{Mol}}$ to the molecular state is 
\begin{equation}\renewcommand\theequation{A\arabic{equation}}\label{eq:Mol_coupling}
\widetilde\Omega_{\mathrm{Mol}} = \sum_{ij}f^\nu_{ij}\widetilde\Omega_{ij}\approx f^\nu\sum_{ij}c_{ij}\widetilde\Omega_{ij},
\end{equation}
where $f^\nu_{ij} = \int\Phi_\nu^{\,*}(R)c^*_{ij}(R)\Phi_{g}(R){\rm d}R$ is the Franck-Condon integral accounting for the overlap of the relative nuclear wave function for each of the contributing pair states $|r_i r_j \rangle$. Here, $\Phi_\nu(R)$ is the vibrational wave function of the molecular state and $\Phi_{g}(R)$ is the relative wave function before PA, in our case given by the Wannier state in the optical lattice.
If the electronic state decomposition of the molecule remains roughly constant over the extension of the vibrational wave function, only a single Franck-Condon integral $f^\nu$ remains.
In this work, this is a good approximation since all hole-hole correlations were measured at the lowest vibrational state.
For higher vibrational modes with larger spatial extension, the changing character of the coefficients becomes important, as discussed in~\cite{Macrodimers_singleatoms_2019}. 
\subsection*{5. \textbf{Inversion symmetry conservation}}\label{section:inv_sym}
Because the Rydberg interaction Hamiltonian commutes with the inversion operator, our macrodimers have a well defined parity, with $p = \pm 1$ for gerade$\,$(ungerade) states~\cite{Weber2017,Samboy2011,Samboy_thesis}. Using the symmetry properties of orbital angular momentum states, symmetrization of pair states $| r_i r_j \rangle$ with respect to inversion yields
\begin{equation}\renewcommand\theequation{A\arabic{equation}}\label{eq:Inv_exp}
|r_{i}r_{j};{g/u}\rangle \propto |r_{i}r_{j}\rangle -p(-1)^{L_i+L_j}|r_{j}r_{i}\rangle,
\end{equation}
with $p=\pm 1$ for gerade~(ungerade) states. For molecular states $|\Psi_\mathrm{Mol};\Omega^{\pm}_{g/u}\rangle = \sum_{i,j}c_{ij}|r_{i}r_{j}\rangle$, this fixes the relation between $c_{ij}$ and $c_{ji}$. Since dipole matrix elements vanish for $\Delta L \neq \pm 1$, the ground state atoms with $L=0$ can only couple states $|r_ir_j\rangle$ with $L_{i/j}=1$, yielding $c_{ij}$ = $\pm c_{ji}$ for coupled ungerade\,(gerade) pair states. From Eq.~(\ref{eq:decomp}), one finds that only for an initial state which has the same inversion symmetry as the molecular state, both added coupling terms $\widetilde\Omega_{ij}$ and $\widetilde\Omega_{ji}$ in Eq.~(\ref{eq:Mol_coupling}) will constructively interfere, while they will cancel each other otherwise. 
\subsection*{6. \textbf{Excitation rates}}
Pairs of Rydberg atoms at small distances autoionize on very fast timescales~\cite{hahn_ionization_2000,robicheaux_ionization_2005}. However, since macrodimers are much larger than the Rydberg orbit, their lifetime is expected to be limited by the decay rates of the individual Rydberg states admixed to the molecule~\cite{schwettmann_analysis_2007,Boisseau2002}. Hence, the decay rate can be calculated from
\begin{equation}\renewcommand\theequation{A\arabic{equation}}\label{eq:Mol_decay}
\gamma_{\mathrm{Mol}} = \sum_{i,j}|c_{i j}|^2(\gamma_{i} + \gamma_{j}),
\end{equation} 
where the single-atom decay rates $\gamma_{i/j}$ include transitions to the ground state as well as room temperature black-body rates to neighboring Rydberg states.


For the $1_u$, $0^-_u$ and $0^+_g$ potentials blue-detuned from the $36P_{1/2}$ resonance studied here, we expect lifetimes of $20.3$, $19.7$ and $20.3\,\mu$s.
We drive the molecular excitation in the incoherent regime, where the scattering rate for a single atom pair can be estimated using
\begin{equation}\renewcommand\theequation{A\arabic{equation}}\label{eq:Mol_scattering}
\gamma_{\mathrm{th}} \approx \frac{\widetilde\Omega_{\mathrm{Mol}}^{2}}{\gamma_{\mathrm{Mol}}}.
\end{equation} 
While the discussion so far was limited to the functional dependence of the microscopic scattering rates on $\beta$,$\vec{\varepsilon}$ and $m_F$, the absolute numbers of the measured scattering rates can also be compared with theory.
All correlation measurements were performed in the low excitation limit with a small mean number of excited molecules $ N_{\mathrm{Mol}} $ where saturation effects are small. 
In that regime, the observed PA rate can be estimated by $\gamma_{\mathrm{exp}} \approx g^{(2)}(1,\pm 1)/t_{\mathrm{uv}}$.
The resulting measured and calculated scattering rates based on Eq.~(\ref{eq:Mol_coupling}), Eq.~(\ref{eq:Mol_decay}) and Eq.~(\ref{eq:Mol_scattering}) are shown in table~\ref{table:Table:3}. 
In agreement with the calculation, we find that the $0^-_u$ potential features the weakest PA rates, related to the large intermediate state detuning close to the center between both fine structure levels $36P_{1/2}$ and $36P_{3/2}$. 
The strongest PA rate was observed for the association of $1_u$ molecules starting from $m_F=-2$, with $\sigma^+$-polarized light.
While the angular dependence presented in Fig.~\ref{fig:2}\,(c), Fig.~\ref{supfig:2}\,(b) and Fig.~\ref{supfig:3}\,(a) shows almost perfect agreement with theory, absolute scattering rates have a higher uncertainty.
This is related with the strong scaling of $\gamma_{\mathrm{th}}$ with the measured reference single-photon Rabi coupling $\widetilde\Omega_{\mathrm{ref}}$.
\section*{Appendix B: \textbf{Zeeman and hyperfine interaction}}\label{B}
\renewcommand{\theequation}{B\arabic{equation}}
\stepcounter{myequation}
The coupling of the molecular state with a magnetic field and the nuclear angular momenta perturbs the Hamiltonian $\hat{H}_0$ of the two isolated atoms and their electrostatic interaction by
\begin{equation}\renewcommand\theequation{B\arabic{equation}}\label{eq:H_1}
\begin{split}
\hat{H}_1 & = \hat{H}^{(1)}_{B}\otimes\mathds{1}^{(2)}+\mathds{1}^{(1)}\otimes\hat{H}^{(2)}_{B} \\ & +\hat{H}^{(1)}_{\mathrm{hfs}}\otimes\mathds{1}^{(2)}+\mathds{1}^{(1)}\otimes\hat{H}^{(2)}_{\mathrm{hfs}},
\end{split}
\end{equation}
with single particle operators
\begin{equation}\renewcommand\theequation{B\arabic{equation}}\label{eq:H_B}
\hat{H}_{B} = \mu_B\left(g_S\hat{\mathbf{S}}+g_L\hat{\mathbf{L}}\right)\cdot\mathbf{B}
\end{equation}
and
\begin{equation}\renewcommand\theequation{B\arabic{equation}}\label{eq:H_hfs}
\hat{H}_{\mathrm{hfs}} =  \sum_{i}A^{r_i}_{\mathrm{hfs}}|r_i \rangle \langle r_i|\hat{\mathbf{I}}\cdot\hat{\mathbf{J}}.
\end{equation}
Here, $\hat{\mathbf{L}}$ and $\hat{\mathbf{S}}$ are orbital angular momentum and electronic spin operators, $g_S$ and $g_L$ the corresponding Landé factors, $\hat{\mathbf{J}}=\hat{\mathbf{L}}+\hat{\mathbf{S}}$ is the total angular momentum of the electron and $\hat{\mathbf{I}}$ the nuclear spin operator. The Zeeman interaction of both isolated nuclei were neglected because of the small g-factor $g_I = 9.95 \times 10^{-4}$~\cite{Arimondo_RevModPhys77}.
The contributing hyperfine constants $A^{r_i}_{\mathrm{hfs}}$ were determined based on measurements at lower principle quantum numbers and known quantum defects, see table~\ref{table:table_qdefects}. We expect $A^{36S_{1/2}}_{\mathrm{hfs}} \approx 487$\,kHz, $A^{36P_{1/2}}_{\mathrm{hfs}} \approx 132$\,kHz and $A^{36P_{3/2}}_{\mathrm{hfs}} \approx 28$\,kHz.
\begin{center}
\begin{table}
\begin{tabular}{ l  c  c  r }
 \hline
  \hline
 \T\B    $nL_{J}$ & $nS_{1/2}$ & $nP_{1/2}$ & $nP_{3/2}$   \\  
 \hline 
 \T$\delta_{0}$ & 3.131\,\cite{Mack2011} & 2.6545\,\cite{LiGallagher_03} & 2.6415\,\cite{LiGallagher_03} \\ 
 \I$\delta_{2}$ & 0.179\,\cite{LiGallagher_03}  & 0.290\,\cite{LiGallagher_03} & 0.295\,\cite{LiGallagher_03}\\ 
 \B\I$A^{\mathrm{ref}}_{\mathrm{hfs}}\,(\mathrm{MHz})$ &  2.14\,($28S_{1/2}$)\,\cite{LiGallagher_03}  & 59.9\,($7P_{1/2}$)\,\cite{feiertag_hyperfine_1973} & 4.05\,($9P_{3/2})$\,\cite{Arimondo_RevModPhys77} \\
  \hline
   \hline
 \end{tabular}
  \caption{\label{table:table_qdefects}Quantum defects and hyperfine constants. The hyperfine constants $A^{r_i}_{\mathrm{hfs}} \propto \left( 1/n^{*}\right) ^3$ are calculated using $n^* = n - \delta (n,L,J)$, $\delta (n,L,J) \approx \delta_0+\left(\frac{\delta_{0}}{n-\delta_{2}}\right)^2$ and literature values $A^{\mathrm{ref}}_{\mathrm{hfs}}$ as references. Contributions from $nD$-states were also included but influenced the value of $A_{\mathrm{eff}}$ by less than 1$\%$, states with $L > 2$ were neglected.}  
\end{table}
\end{center}
For the energetically isolated $1_u$ potential presented in Fig.~\ref{fig:3}, only the two states $|\Psi_{\mathrm{Mol}};\pm 1_u\rangle = \sum_{i,j}c^{\pm 1}_{ij}|r_{i}r_{j}\rangle$ with $\Omega = \pm 1$ are relevant.  
Hence, energy shifts can be calculated within the subspace $ \{|\Psi_{\mathrm{Mol}};\pm 1_u\rangle \otimes |m_{I_1},m_{I_2}\rangle\}$, with $m_{I_{1(2)}}$ the nuclear spin orientation of both atoms.
Because $\hat{\mathbf{S}}$ and $\hat{\mathbf{L}}$ cannot change $\Omega$ by $2$, none of the 32 relevant states can be coupled by Eq.~\ref{eq:H_B}.
Writing $\hat{\mathbf{I}}\cdot\hat{\mathbf{J}} = \frac{1}{2}(\hat{I}^-\hat{J}^+ + \hat{I}^+\hat{J}^-) + \hat{I}^z\hat{J}^z$ with $\hat{J}^\pm = \hat{J}^x\pm i\hat{J}^y$ and $\hat{I}^\pm = \hat{I}^x\pm i\hat{I}^y$, one finds the same result for Eq.~\ref{eq:H_hfs}. 
Consequently, all angular momentum operators can be replaced by their $z-$components and magnetic field components $B_{x}$ and $B_{y}$ can be neglected. The presented energy shifts were obtained by
\begin{equation}\renewcommand\theequation{B\arabic{equation}}\label{eq:Zeeman_Hf}
\Delta E^\Omega_{m_{I_{1}},m_{I_{2}}} = g_{\mathrm{eff}}\mu_{B}\Omega B_z + A_{\mathrm{eff}}(m_{I_1}+m_{I_2})\frac{\Omega}{2}.
\end{equation}
Using $g_S\hat{S}_z + g_L\hat{L}_z = g_L\hat{J}_z + (g_S-g_L)\hat{S}_z$ with $g_L = 1$ and $g_S \approx 2$, the effective molecular g-factor writes 
\begin{equation}\renewcommand\theequation{B\arabic{equation}}
\begin{split}
g_{\mathrm{eff}} = &\hspace{4pt} |\pm 1 + \langle \Psi_{\mathrm{Mol}};\pm 1_u |\hat{S}_z^{(1)}\otimes\mathds{1}^{(2)}|\Psi_{\mathrm{Mol}}; \pm 1_u \rangle \\ & + \langle \Psi_{\mathrm{Mol}};\pm 1_u |\mathds{1}^{(1)}\otimes\hat{S}_z^{(2)}| \Psi_{\mathrm{Mol}};\pm 1_u \rangle | ,
\end{split}
\end{equation}
yielding $g_{\mathrm{eff}} \approx 0.66$ after expanding the contributing asymptotic states $|r_i,r_j\rangle$ into uncoupled spin and orbital angular momenta.
The effective hyperfine interaction was calculated as $A_{\mathrm{eff}} = \sum_{i,j}|c^{\pm 1}_{ij}|^2(A^{r_i}_{\mathrm{hfs}}m^{\prime}_{Ji}+ A^{r_j}_{\mathrm{hfs}}m^{\prime}_{Jj}) \approx 127\,$kHz. Because the molecular bond lengths and the lattice spacing are larger than the contributing Rydberg orbits ($\approx 130$\,nm), we only consider the coupling of the Rydberg states to the nucleus they are bound to.
\begin{figure}[htb]
  \centering
  \includegraphics[width=1.0\columnwidth]{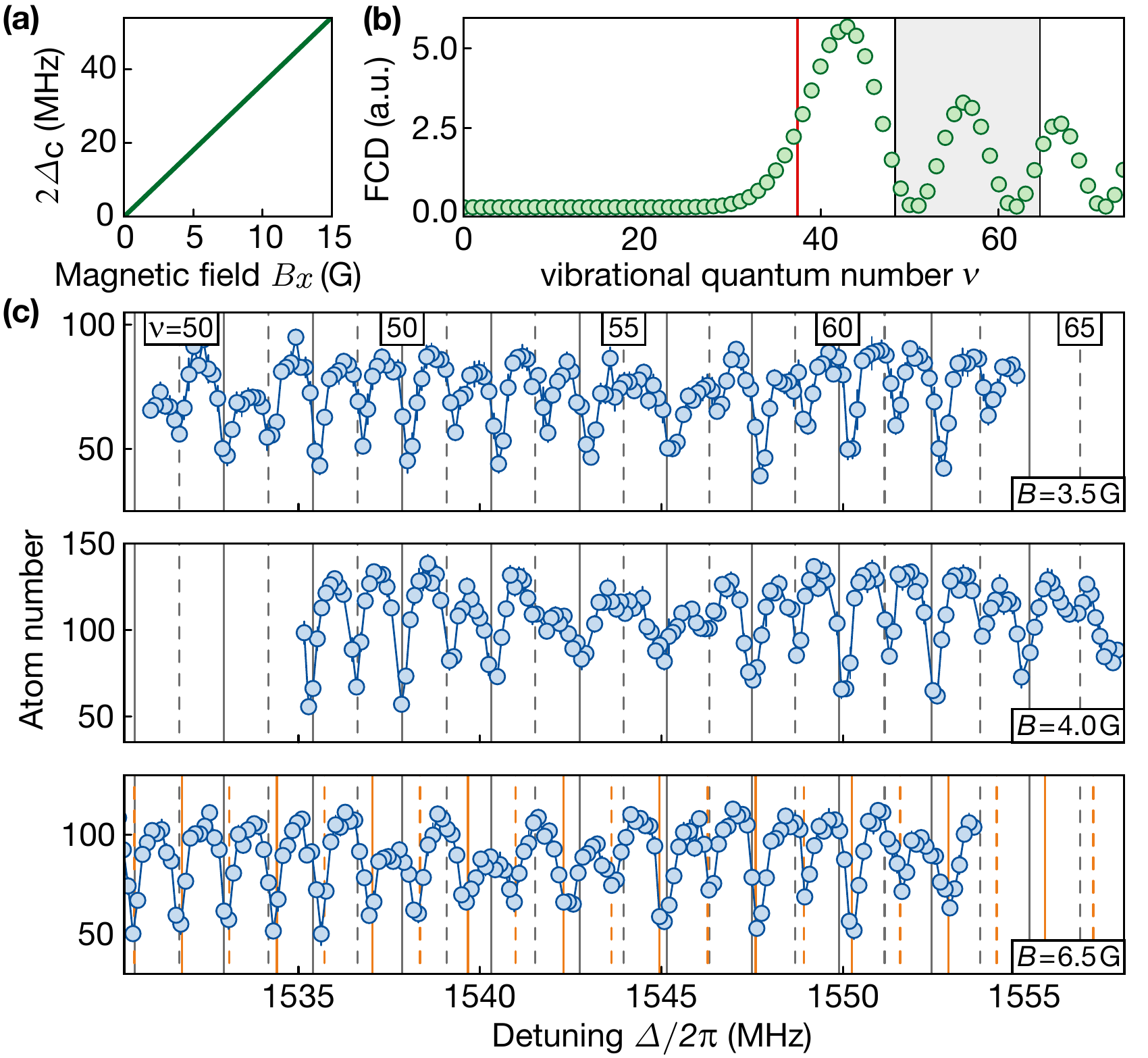}%
  \caption{ \label{supfig:4}
Further details on the predissociation. (a) Calculated magnetic field dependence of the gap between both potentials. (b) Calculated Franck-Condon density coupling bound states to continuum modes at same energy in arbitrary units (a.u.). The calculated position of the crossing and the measured frequency region are indicated as a red line and a grey shaded area. (c) Further spectroscopic results starting from $m_F = 0$.  For $B = 3.5\,$G and $B = 4.0\,$G, grey lines denote the calculated resonance frequencies for the $0^-_u$ potential at zero field. For $B = 6.5\,$G, the calculations in the combined potential were included (orange). Because of an overall energy shift observed at $B = 6.5\,$G, both sets of resonances were manually shifted to agree with the lowest vibrational resonance of the spectrum. All error bars on the data points denote one s.e.m.}
\end{figure}
To further verify the presence of the hyperfine coupling, we also measured the magnetic field splitting for the corresponding $1_u$ potential at the blue-detuned side of $31P_{1/2}$, see Fig.~\ref{supfig:3}\,(b). 
The reduced bond length close to the lattice constant now leads to a correlation signal peaking at a distance of $a_{\mathrm{lat}}$. This requires a magnetic field applied along the lattice direction in order to keep the conditions identical to Fig.~\ref{fig:3}\,(b).
While $g_{\mathrm{eff}}\approx 0.63$ remains almost the same, the hyperfine interaction $A_{\mathrm{eff}}\approx 216\,$kHz is larger. 
This leads to an even more significant deviation from a calculation neglecting the hyperfine interaction.
Starting from $m_F=0$, we also reproducibly observed that the $\Omega = - 1$ resonance appears stronger than the $\Omega = + 1$ resonance, which might be due to a slightly imperfect UV polarization. 
\section*{Appendix c: \textbf{Breaking molecular symmetries}}\label{C}
\renewcommand{\theequation}{C\arabic{equation}}
\stepcounter{myequation}
For the crossing potentials $V_{1_u}(R)$ and $V_{0^-_u}(R)$ discussed in Fig.~\ref{fig:4}, the relevant molecular states are $|\Psi_{\mathrm{Mol}};\pm 1_u\rangle$ and $|\Psi_{\mathrm{Mol}};0^-_u\rangle$.
We define states $|\Psi_{\mathrm{Mol}};S_u\rangle = 1/\sqrt{2}(|\Psi_{\mathrm{Mol}};+1_u\rangle + |\Psi_{\mathrm{Mol}};-1_u\rangle)$ and $|\Psi_{\mathrm{Mol}};A_u\rangle = 1/\sqrt{2}(|\Psi_{\mathrm{Mol}};+ 1_u\rangle - |\Psi_{\mathrm{Mol}};-1_u\rangle)$. 
Neglecting the hyperfine part, Eq.~\ref{eq:H_1} reduces to $\hat{H}_1  = \hat{H}^{(1)}_{B}\otimes\mathds{1}^{(2)}+\mathds{1}^{(1)}\otimes\hat{H}^{(2)}_{B}$.
For $\mathbf{B}\perp\mathbf{R}$, the single particle Hamiltonians can be expressed as 
\begin{equation}\renewcommand\theequation{C\arabic{equation}}
\hat{H}_{B} = \mu_B\left(g_S\hat{S}_x+g_L\hat{L}_x\right) B_x.
\end{equation}
The magnetic field induces a coupling $\Delta_C(R) = \langle \Psi{_\mathrm{Mol}};0^-_u | \hat{H}_1 |\Psi_{\mathrm{Mol}};S_u \rangle $. 
Because $\langle\Psi_{\mathrm{Mol}};+1_u|\hat{H}_1|\Psi{_\mathrm{Mol}};0^-_u \rangle = \langle\Psi_{\mathrm{Mol}};-1_u|\hat{H}_1|\Psi{_\mathrm{Mol}};0^-_u \rangle$, the state $|\Psi_\mathrm{Mol};A_u\rangle$ remains uncoupled, leaving one of the two crossing $1_u$ potential curves unchanged. 
Again, the molecular state decompositions depend on $R$.  
At the crossing point $R_c$ of both potentials, the gap size is $2 \Delta_C(R_c)$, as shown in Fig.~\ref{supfig:4}\,(a). 
The new combined potentials shown in  in Fig.~\ref{fig:4}\,(b) are obtained by diagonalizing
\begin{equation}\renewcommand\theequation{C\arabic{equation}}
\hat{H} = \begin{pmatrix}
V_{1_u}(R) & \Delta_C(R) \\ \Delta_C(R) & V_{0^-_u}(R)
\end{pmatrix}.
\end{equation}
The coupling of the bound states to the continuum does not only depend on the electronic coupling $\Delta_C(R)$ but also on the overlap with the nuclear motion. 
In our case, where one potential is a repulsive potential well, a Franck-Condon density (FCD) quantifies the coupling strength between a vibrational state $\Phi_\nu(R)$ to nearby continuum modes.
An estimation of the FCD by the overlap integral of the vibrational bound states with the continuum states at same energy is shown in Fig.~\ref{supfig:4}\,(b). 
We find that the FCD vanishes for vibrational states energetically below the crossing, while it oscillates with $\nu$ above. 
This is consistent with previous studies on predissociation~\cite{vigue_j._natural_1981_2} and can be understood by comparing the first lobe of the continuum states with the vibrational states $\Phi_\nu (R)$:
If it coincides with the first lobe of $\Phi_\nu  (R)$, the FCD reaches a maximum.
It becomes very small if it matches with the first zero crossing of $\Phi_\nu  (R)$ and increases again matching with the second lobe of $\Phi_\nu (R)$ and so on.

Further spectroscopic data for $B=3.5\,$G, $B=4.0\,$G and $B=6.5\,$G, again for an initial state $m_F=0$, are shown in Fig.~\ref{supfig:4}\,(c).
For $B=3.5\,$G, we start to see first indications of the gap at $\nu\approx 55$. Additionally, it seems that another region of broadened vibrational resonances occurs for $\nu \lesssim 50$.
For $B=4.0\,$G, perturbations are clearly visible at $\nu\approx 55$ and also seem to be present for $\nu \gtrsim 66$.
For both datasets, the vibrational resonances are still in good agreement with the calculated eigenenmodes in the $0^-_u$ potential at zero field.
For $B=6.5\,$G, the perturbed region moved to slightly lower energies and is significantly weaker.
Also, the spectrum is now shifted and the experimentally observed vibrational spacing $\Delta\nu_{\mathrm{exp}} = 2.66(1)\,$MHz is in agreement with the calculated result $\Delta\nu_{\mathrm{comb}}=2.66(2)\,$MHz in the combined potential rather than the $\Delta\nu_{\mathrm{0^-_u}}=2.45(1)\,$MHz in the $0^-_u$ potential. In all cases, $\Delta\nu_{\mathrm{exp}}$ was obtained by fitting a vibrational series with varying frequency spacing to the data. Uncertainties on the calculated values $\Delta\nu_{\mathrm{0^-_u}}$ and $\Delta\nu_{\mathrm{comb}}$ account for the anharmonicity of both molecular potentials in the frequency range of the spectroscopies performed above the potential crossing.
For $B=6.5\,$G, both sets of eigenenergies were manually shifted to agree with the lowest vibrational resonance of the spectroscopy. 
\section*{Appendix d: \textbf{Methods}}\label{D}
All macrodimer potentials and wave functions were obtained by the pair interaction program~\cite{Weber2017}. 
Agreement between measured and calculated vibrational resonances required $10\,000-15\,000$ basis states, within an energy band of around $500\,$GHz. 
Perturbations by magnetic fields were calculated using the molecular states obtained at zero field and the presented formalism.
Rydberg lifetimes contributing to the molecular decay rates were calculated with the ARC package~\cite{Sibalic2017}.
The rotational constant was calculated as $B_{\mathrm{rot}} = \hbar^2/(4 \mu a^2_{\mathrm{lat}})$~\cite{Demtroeder_rotation_2007}, with $\mu$ the reduced mass of both $^{87}\mathrm{Rb}$ atoms.

All detunings $\Delta$ are defined relative to the center of the Zeeman split states in the ground state and the Rydberg state. 
The Rydberg resonances were drifting by up to $500$\,kHz on timescales of roughly a day, which required us to track the resonance position. 
We always chose a lattice depth of $V_0 \approx 40\,E_{\mathrm{rec}}$ for both lattices in the atomic plane and $V_0 \approx 80\,E_{\mathrm{rec}}$ perpendicular to it. Here, $E_{\mathrm{rec}}$ is the recoil energy of the optical lattice~\cite{BlochZwergerDalibard}.
Since the Wannier state in the atomic limit is only weakly dependent on the lattice depth, this is not a critical parameter as long as one stays deep in the Mott insulating regime~\cite{Macrodimers_singleatoms_2019,BlochZwergerDalibard}.
In order to evaluate $\gamma_{\mathrm{th}}$ by Eq.~\ref{eq:Mol_scattering}, we measure the single atom Rabi frequency $\widetilde \Omega_{\mathrm{ref}}$ coupling the ground state $|m_F=-2 \rangle$ to the Rydberg state $|36P_{1/2},m^\prime_J=+1/2\rangle$ with linear UV polarization perpendicular to the chosen magnetic field. 
We obtain $\widetilde \Omega_{\mathrm{ref}}$ by probing the AC-Stark shift at a detuning $\Delta$ from the Rydberg resonance with a microwave field driving both ground state hyperfine levels $F=1$ and $F=2$. 
For the spectroscopic data, the frequency was swept during the UV pulse to cover the full range between neighboring data points.
The data points always represent the mean of roughly 10 shots.
For each of the correlation measurements, we took at least 200 images under the same conditions. 
We also studied the ejection efficiency of Rydberg atoms from the lattice by measuring the $F=1$ fraction of atoms after a Rydberg excitation pulse starting from $F=2$ in a configuration where excited atoms are expected to also decay to $F=1$. 
We did not find an increase of atoms in $F=1$ and conclude that retrapping does not occur. 
Experimental runs with high atom number fluctuations due to an imperfect preparation of the ground state atom array were excluded. 
For the results presented in Fig.~\ref{fig:2}\,(c) and Fig.~\ref{supfig:2}\,(b), we rotated the magnetic field with an absolute value of $B = 1.0\,$G in five steps from $\beta = 0^\circ$ to $\beta = 90^\circ$ to get the characteristic curves. 
The symmetry allows us to take the data in this region only.
In Fig.~\ref{fig:3}\,(b), we fit a sum of three Gaussians and use the result for the central line as a reference to indicate the calculated Zeeman and hyperfine shifts.

\bibliography{Molecular_symmetries_revtex}

\end{document}